\documentclass[epj]{webofc}
\usepackage[varg]{txfonts}   
\usepackage[colorlinks=true,citecolor=blue]{hyperref}
\woctitle{RICAP--14 - Roma International Conference on Astro-Particle Physics }
\begin{document}
\title{The ASTRI mini-array within the future Cherenkov Telescope Array}
%
%

\author{Stefano Vercellone\inst{1}\fnsep\thanks{\email{stefano.vercellone@iasf-palermo.inaf.it}} 
	\and
        for The ASTRI Collaboration\inst{2} \and
        The CTA Consortium\inst{3}
}

\institute{INAF/IASF Palermo, Via U. La Malfa 153, 90146 Palermo, Italy 
\and
           http://www.brera.inaf.it/astri/ 
\and
           https://portal.cta-observatory.org
          }

\abstract{%
The Cherenkov Telescope Array (CTA) is a large collaborative effort
aimed at the design and operation of an observatory dedicated to
very high-energy gamma-ray astrophysics in the energy range from a few tens of GeV to above 100\,TeV,
which will yield about an order of magnitude improvement in sensitivity with 
respect to the current major arrays (H.E.S.S., MAGIC, and VERITAS).
Within this framework, the Italian National Institute for Astrophysics is leading
the ASTRI project, whose main goals are the design and installation on Mt. Etna
(Sicily) of an end-to-end dual-mirror prototype of the CTA small size telescope (SST) 
and the installation at the CTA Southern site of a dual-mirror SST mini-array composed 
of nine units with a relative distance of about 300\,m.
The innovative dual-mirror Schwarzschild-Couder optical solution adopted for the 
ASTRI Project allows us to substantially reduce the telescope plate-scale and, therefore,
to adopt silicon photo-multipliers as light detectors.
The ASTRI mini-array is a wider international effort. The mini-array,
sensitive in the energy range 1--100 TeV and beyond with an angular resolution 
of a few arcmin and an energy resolution of about 10--15\%, is well suited to study 
relatively bright sources (a few $\times 10^{-12}$\,erg\,cm$^{-2}$s$^{-1}$ at 10\,TeV) at very high energy.
Prominent sources such as extreme blazars, nearby well-known BL Lac objects, Galactic pulsar 
wind nebulae, supernovae remnants, micro-quasars, and the Galactic Center can be observed 
in a previously unexplored energy range. The ASTRI mini-array will extend the 
current IACTs sensitivity well above a few tens of TeV and, at the same time, will allow us to
compare our results on a few selected targets with those of current (HAWC) and future
high-altitude extensive air-shower detectors.
}
\maketitle
%
\section{Introduction}
\label{intro}
The very high-energy (VHE) portion of the electromagnetic spectrum (above $\approx 100$\,GeV)
is currently being investigated by means of ground-based imaging atmospheric Cherenkov telescopes
(IACTs, see~\cite{2009ARA&A..47..523H} for a recent review).
In order to dramatically boost the current IACT performance and to widen the VHE science, a new 
Cherenkov telescope array (CTA) has been proposed, as described in~\cite{2011ExA....32..193A} and more
recently in~\cite{2013APh....43....3A}.
The wide energy range covered by the CTA (from a few tens of GeV to above 100\,TeV) requires
different kinds of telescopes. Four large size telescopes (LSTs, D$\sim23$\,m) will be placed at the center of the array,
to lower the energy threshold down to a few tens of GeV. A few tens of medium size telescopes 
(MSTs, D$\sim12$\,m, SCTs, D$\sim9.5$\,m) will cover 
approximately 1\,km$^{2}$, to improve by a factor of ten the sensitivity in the energy range 0.1--10\,TeV.
Finally, 70 small size telescopes 
(SSTs, primary mirror D$\sim4$\,m, Aeff$\sim5-10$\,m$^{2}$) covering about 10\,km$^{2}$
will enhance Galactic plane source studies in the energy range beyond 100 TeV.
A detailed review of the CTA project is given in~\cite{JohnCarr}.
\begin{figure}
\centering
\includegraphics[width=6.9cm,clip]{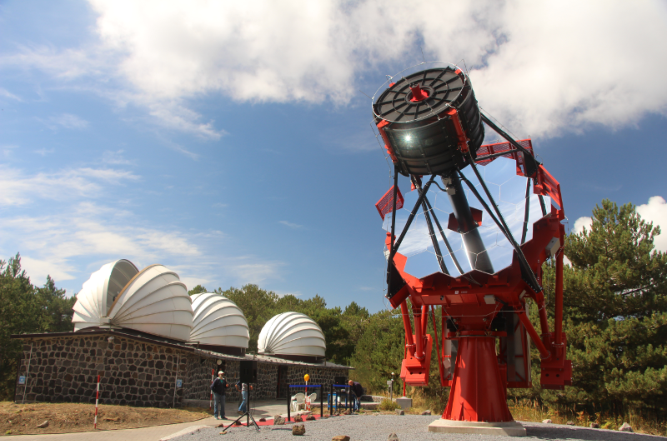}
\includegraphics[width=6.9cm,clip]{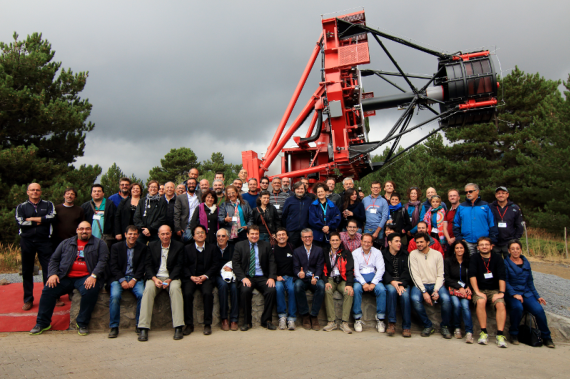}
\caption{{\it Left panel}: the ASTRI SST-2M prototype at the Serra La Nave observing station a few hours before the
official inauguration.  {\it Right panel}: the ASTRI Collaboration in front of the telescope.}
\label{fig:fig1}      
\end{figure}
%

\section{The ASTRI Project and its end-to-end prototype}
\label{astri}
Within the CTA framework, the Italian National Institute for Astrophysics (INAF) is leading  the 
``Astrofisica con Specchi a Tecnologia Replicante Italiana'' (ASTRI)
Flagship Project~\cite{2013arXiv1307.4962P} of the Ministry of
Education, University and Research.
Primarily, INAF has designed and developed an end-to-end prototype of the CTA small-size telescope 
in a dual-mirror configuration (SST-2M). This prototype is currently being tested under field conditions at the 
INAF ``M.C. Fraca\-sto\-ro'' observing station in Serra La Nave (Mount Etna, Sicily).
The ASTRI SST-2M prototype was recently inaugurated during the CTA Consortium Meeting
in September 2014. Fig.~\ref{fig:fig1} shows the prototype in front of the domes (left panel) and behind
the ASTRI Collaboration a few moments after the inauguration (right panel).

A new dual-mirror, Schwarzchild-Couder (SC) aplanatic design has been
proposed~\cite{2007APh....28...10V}.
In the SC telescope, the focal plane is located in-between two aspherical mirrors, close to the secondary mirror.
No Cherenkov telescope has adopted this optical system before.
The dual-mirror optical system will reduce the dimension, the weight, and the cost of the camera
at the focal plane of the telescope, and will obtain a more compact and stiffer mechanical structure, and
 an optimal imaging resolution across a wide field of view.
Moreover, due to the reduced plate-scale, silicon-based photo-multipliers (SiPMs) can be adopted as light detectors.

The ASTRI SST-2M prototype adopts a segmented 4.3\,m primary mirror (M1) composed of 18 facets, a monolithic 1.8\,m
secondary mirror (M2, with a radius of curvature of 2.2\,m), a focal length F=2.15\,m, a field of view FoV$\sim9.6^{\circ}$,
for a ratio F/D$_{1}$=0.5. The mirror manufacturing process is the ``glass cold shaping'' technique, specifically developed 
by INAF for Cherenkov mirrors~\cite{2014SPIE.9145E..0MC, 2014SPIE.9151E..0MC}.
The curved focal plane ($\sim 1$\,m of radius of curvature) hosts 1984 logical pixels
(6.2\,mm\,$\times$\,6.2\,mm, $0.17^{\circ})$. The current photo-sensors are Hamamatsu S11828-3344M 
silicon-based photo-multipliers, but other sensors are also being tested~\cite{6825899}. 
The ASTRI camera~\cite{2014SPIE.9147E..0DC} is extremely compact ($\sim$\,50\,cm\,$\times$\,50\,cm\,$\times$\,50\,cm) and
light ($\sim$\,50\,kg). Contrary to other CTA telescopes adopting a signal-sampler front-end electronics (FEE), the ASTRI camera 
adopts as FEE the CITIROC, a customized version of the EASIROC~\cite{Callier20121569} ASIC signal-shaper manufactured 
by Omega\footnote{\href{http://omega.in2p3.fr/}{\tt http://omega.in2p3.fr/}; manufactured under INAF intellectual property.}.
%

The ASTRI SST--2M is mainly a technological prototype. Nevertheless, after a thorough commissioning phase,
it will perform scientific observations of the Crab Nebula, Mrk~421, and Mrk~501. These observations will allow us
to perform the science verification phase, in order to cross-check the prototype performance with our Monte Carlo
simulations. We estimate that, in the maximum sensitivity range (E$\ge 2$\,TeV),
we can detect a flux level of 1~Crab at 5$\sigma$ in a few hours~\cite{2013arXiv1307.5006B}.

\section{The ASTRI mini-array}
\label{miniarray}
\begin{figure}
\centering
\sidecaption
\includegraphics[width=7.6cm,clip]{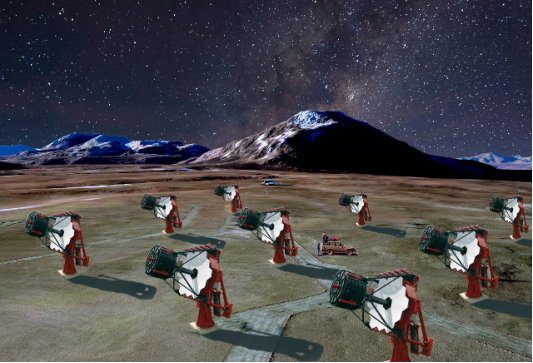}
\caption{Artistic concept (not to scale) of the ASTRI mini-array. The ASTRI mini-array is a collaborative
effort, within the CTA framework, among Italy, Brazil and South Africa.
Two Brazilian engineers are currently working at the
Serra La Nave site.
The ASTRI mini-array is proposed to be placed at the CTA southern site.}
\label{fig:fig2}      
\end{figure}

A remarkable improvement in terms of performance could come from the operation, in late 2016, of a mini-array, 
composed of nine SST-2M telescopes (see Fig.~\ref{fig:fig2} for an artistic concept) to be placed at the final CTA southern site. 
Preliminary Monte Carlo simulations~\cite{ASTRI_MC_OATO_5000_016}     
yield an improvement in sensitivity that, for nine telescopes, could surpass the H.E.S.S. sensitivity
above 10\,TeV, extending up to about 100\,TeV.
The ASTRI mini-array will be able to study in great detail relatively bright 
(a few $\times 10^{-12}$\,erg\,cm$^{-2}$s$^{-1}$ at 10~TeV) sources with an angular resolution of a few 
arcmin and an energy resolution of about 10--15\,\%. 

Thanks to the array approach, it will be possible to verify the wide FoV performance to detect very high
energy showers with the core located at a distance up to 500\,m and to compare the mini-array performance with the Monte
Carlo expectations by means of deep observations of a few selected targets.
Moreover, it will be possible to perform the first CTA science,
with its first solid detections during the first year of operation, as described in~\cite{2013arXiv1307.5671V}.
Prominent sources such as extreme blazars (KUV~00311$-1938$), nearby well-known BL~Lac objects (Mrk~501)
and radio-galaxies (M~87), galactic pulsar wind nebulae (Crab Nebula, Vela-X), supernovae remnants
(Vela-junior, RX~J1713.7$-$3946), as well as the Galactic Center can
be observed in a previously unexplored energy range, in order to investigate the electron acceleration and cooling,
relativistic and non relativistic shocks, the search for cosmic-ray (CR) PeVatrons, the study of the CR propagation,
and the impact of the extragalactic background light on the spectra of the sources.
\begin{figure}
\centering
\includegraphics[width=6.6cm,clip]{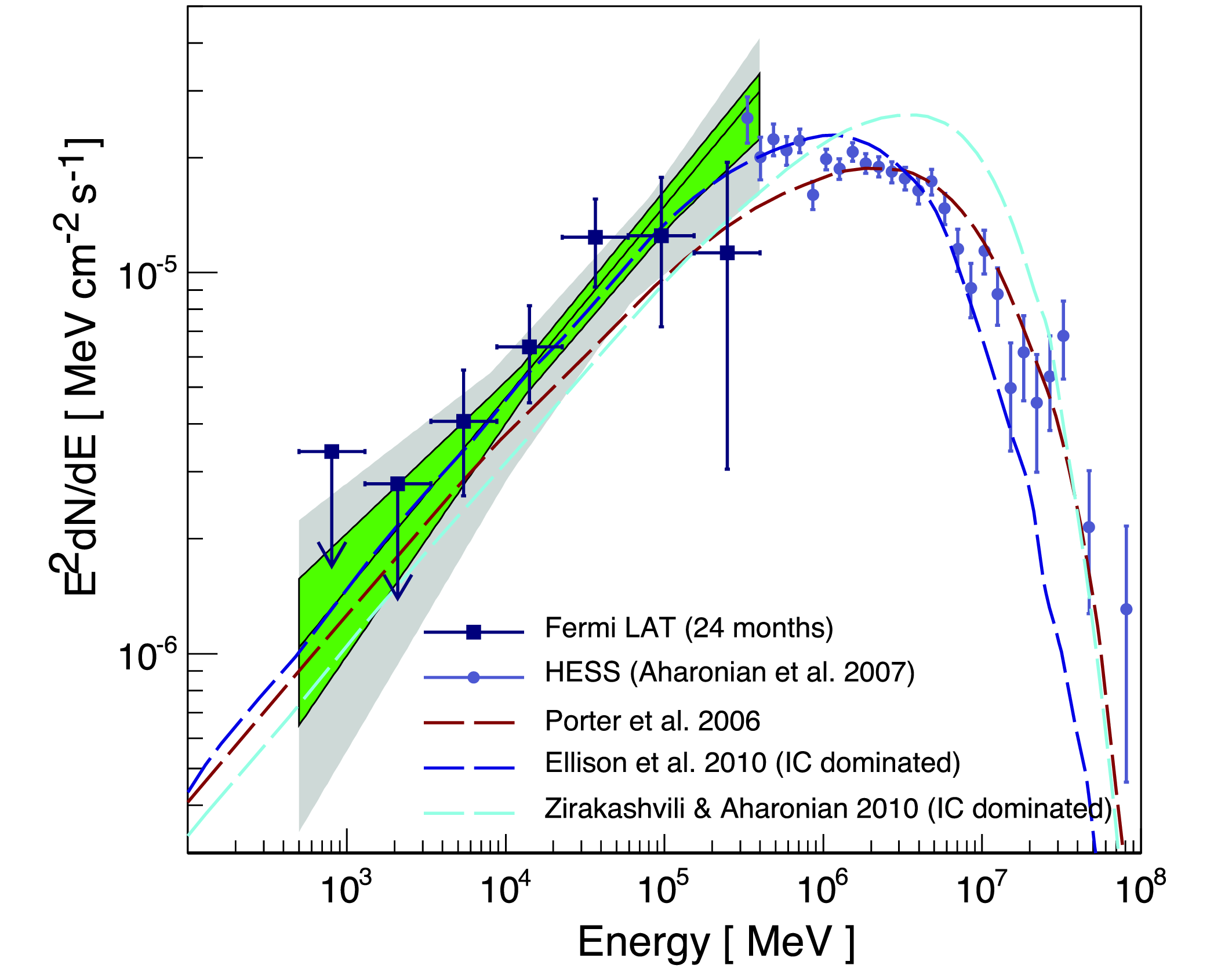}
\includegraphics[width=6.6cm,clip]{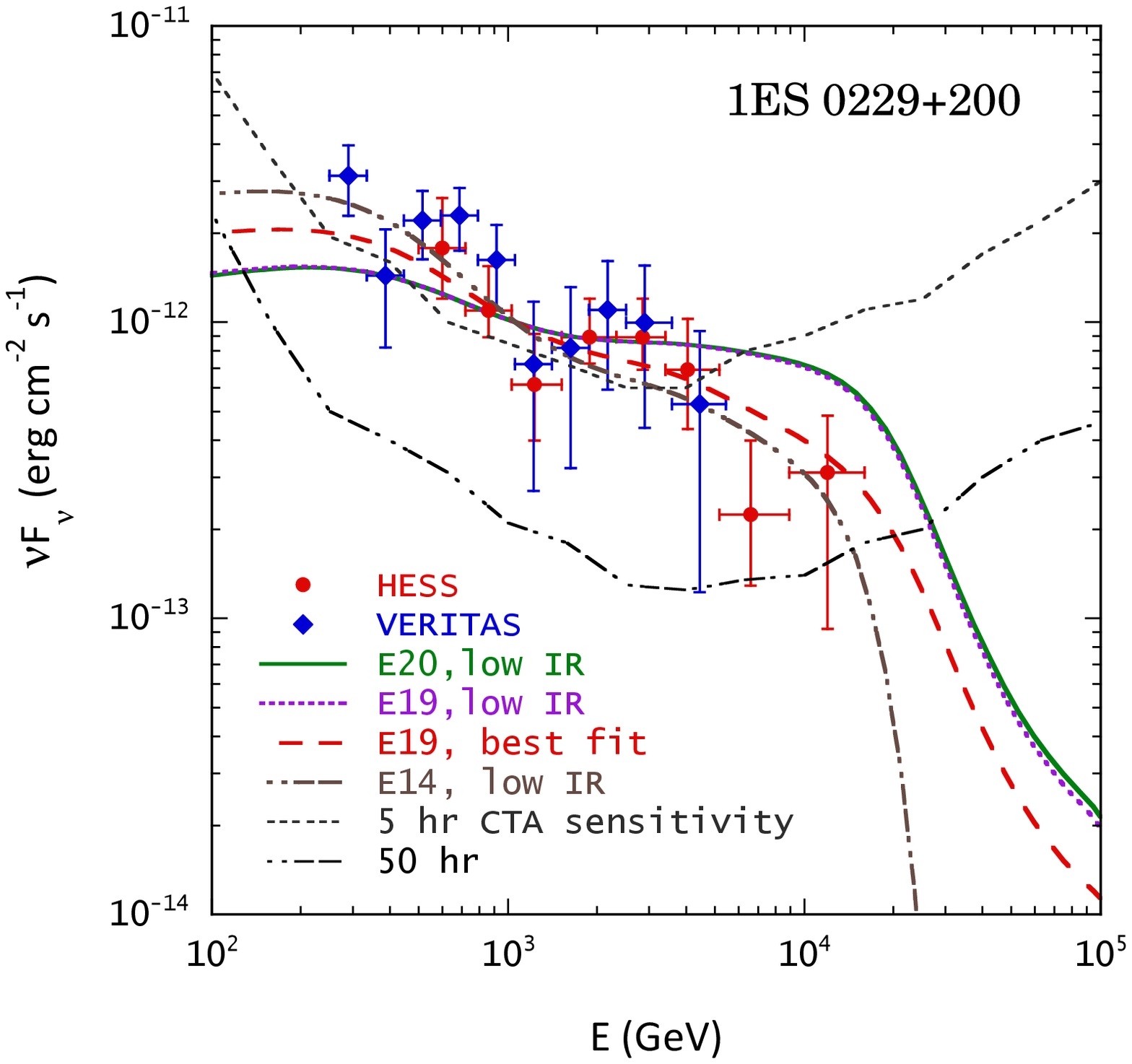}
\caption{{\it Left panel}:  supernova remnant RX~J1713.7$-$3946 (see~\cite{2011ApJ...734...28A} for details). 
We expect for the ASTRI mini-array sensitivity at least comparable to (or slightly better than) the H.E.S.S. 
one above a few TeV. 
{\it Right panel}: spectral energy distribution of the extreme blazar 1ES~0229$+$200 
(see~\cite{2012ApJ...749...63M} for details).}
\label{fig:fig3}      
\end{figure}

On the Galactic plane, one of the best targets is RX~J1713.7$-$3946. Its detection by {\it Fermi}~\cite{2011ApJ...734...28A} 
and the combined study with H.E.S.S. (Fig.\ref{fig:fig3}, left panel) shows that the high-energy and very high-energy (VHE) 
emission could be interpreted in the framework of a leptonic scenario.
The improved and uniform sensitivity (within a few degrees off-axis) and the comparable angular resolution of the ASTRI
mini-array at E$\ge 10$\,TeV with respect to the current 
IACTs could allow us to investigate the VHE emission in the different regions of this source, studying their spectra, and 
to extend the current spectral energy distribution (SED) well above a few tens of TeV, searching for possible spectral cut-offs.
Fig.~\ref{fig:fig3} (right panel) shows the SED of the extreme blazar
1ES~0229$+$200 with superimposed different theoretical SED fits (both hadronic and leptonic) assuming 
different EBL models (see~\cite{2012ApJ...749...63M} for a detailed discussion) and the 5-$\sigma$ differential sensitivity
for 5 and 50\,hr observations with CTA (configuration E, as reported in~\cite{2011ExA....32..193A}).
Because of the uncertainty in EBL models, it is not easy to distinguish between the hadronic and leptonic scenarios at
$\sim$1-10\,TeV energies. At higher energies, however, UHECR-induced cascade emission becomes 
harder than the gamma-ray one. A detection of $\ge 25$\,TeV gamma-rays from 
1ES 0229$+$200 would only be consistent with an hadronic scenario.

\section{Synergies and conclusions}
\label{synergies}
%
The ASTRI mini-array will operate when the other IACTs will still be active. Compared to them,
the ASTRI mini-array will extend the sensitivity up to 100 TeV and beyond, a never-explored energy range by IACTs. 
Moreover, it will benefit from a much larger field of view which will allow us to study in detail
extended sources at energies about a decade higher than what is currently being explored and
to monitor simultaneously a few close-by sources during the same pointing.
Long exposures will be preferred, restricting the number of possible targets, and extending the observations 
also during moon light periods, thanks to the use of a SiPMs-based camera.

The lower imaging energy threshold of current and future extended air-shower (EAS) detectors ($\sim$100\,GeV)  
and the wider energy range of the ASTRI mini-array (beyond 100\,TeV) will allow us a direct comparison 
of scientific data (spectra, light-curves, integral fluxes) of those sources which could be monitored simultaneously,
although on different integration time-scales. Moreover, the high-energy boundary 
of both EAS and the ASTRI mini-array will allow us to study the VHE (E$\ge 10$\,TeV) emission from extended 
source such as SNRs and PWN, and to investigate the presence of spectral cut-offs.

In summary, the ASTRI mini-array could be considered as the first CTA {\it seed}, allowing the entire CTA Consortium
to start seminal studies on both Galactic and extra-galactic sources.

\section*{Acknowledgements}
\label{acknowledgement}
This work was partially supported by the ASTRI ``Flagship Project'' financed by the Italian Ministry of Education, 
University, and Research (MIUR) and led by the Italian National Institute of Astrophysics (INAF). 
We acknowledge partial support by the MIUR Bando PRIN 2009 and TeChe.it 2014 Special Grants. 
We also acknowledge support from the Brazilian Funding Agency FAPESP (Grant 2013/10559-5) 
and from the South African Department of Science and Technology through Funding Agreement 0227/2014 
for the South African Gamma-Ray Astronomy Programme.
We gratefully acknowledge support from the agencies and organizations 
listed in this page: \href{http://www.cta-observatory.org/?q=node/22}{\tt http://www.cta-observatory.org/?q=node/22}.
 %
%
%
%


%
%

\end{document}